\documentclass[twocolumn,showpacs,preprintnumbers,amsmath,amssymb,aps,prb,floatfix,groupedaddress]{revtex4}
\usepackage{graphicx} 
\usepackage{dcolumn} 
\usepackage{bm} 
\usepackage{epsfig}
\usepackage{color}
\usepackage{multirow}

\usepackage{amsmath}

\begin{document}

\title{Linear bands, zero-momentum Weyl semimetal, and topological transition
             in skutterudite-structure pnictides}
\author{V. Pardo}
\affiliation{Departamento de F\'{i}sica Aplicada, Universidade
     Santiago de Compostela, Spain, and
     Department of Physics, University of California Davis}
\author{J. C. Smith}
\affiliation{Department of Physics, University of California Davis}
\author{ and W. E. Pickett}
\affiliation{Department of Physics, University of California Davis}

\begin{abstract}
It was reported earlier [Phys. Rev. Lett. 106, 056401 (2011)] that 
the skutterudite structure compound CoSb$_3$ displays
a unique band structure with a topological transition versus a symmetry-preserving
sublattice (Sb) displacement very near the structural ground state.  The transition
is through a massless Dirac-Weyl semimetal, point Fermi surface phase 
which is unique in that (1) it
appears in a three dimensional crystal, (2) the band critical point occurs
at $k$=0, and (3) linear bands are degenerate with conventional (massive) bands at the critical
point (before inclusion of spin-orbit coupling).  Further interest arises because 
the critical point separates a conventional (trivial)
phase from a topological phase.  In the native cubic
structure this is a zero-gap topological semimetal; we show how
spin-orbit coupling and uniaxial strain converts 
the system to a topological insulator (TI).
We also analyze the origin of the
linear band in this class of materials, which is the characteristic that
makes them potentially useful
in thermoelectric applications or possibly as transparent conductors.
We characterize the formal charge as Co$^{+}$ $d^8$, consistent with the gap, with its
$\bar{3}$ site symmetry, and with its lack of moment. The Sb states are characterized as
$p_x$ (separately, $p_y$) $\sigma$-bonded $Sb_4$ ring states occupied and the
corresponding antibonding states empty. The remaining (locally) $p_z$ orbitals
form molecular orbitals with definite parity centered on the empty $2a$ site
in the skutterudite structure. Eight
such orbitals must be occupied; the one giving the linear band is an odd orbital
singlet $A_{2u}$ at the zone center.  We observe that the provocative linearity of the band within the
gap is a consequence of the aforementioned near-degeneracy, 
which is also responsible for the small
band gap.

\end{abstract}
\maketitle
\date{\today}

\vskip2pc

\section{Motivation and Introduction}
Recent years have seen an explosion of interest in two areas that, while distinct
in themselves, have also found some areas of overlap.  One is the situation where
Dirac-Weyl linear bands emanate from a point Fermi surface (FS), the so-called Dirac point.
While linear bands around symmetry points  are not uncommon in
crystal structures, it is uncommon that such points can determine the Fermi energy
(E$_F$), which is what happens in the celebrated case of graphene. It is necessary
that there is a gap throughout the Brillouin zone except for the touching bands.
Point FSs occur also in the case of conventional (massive) zero-gap semiconductors,\cite{book}
and give rise to properties that have been studied in some detail. The case of
the Dirac-Weyl semimetal (point Fermi surface) systems is quite 
different,\cite{graph1,graph2} leading to a great
deal of new phenomenology. 

The other new and active area relevant to this paper is that of topological insulators
(TIs), in which the Brillouin zone integral of a certain `gauge field' derived from the
$k$-dependence of the periodic part of the Bloch states of the crystal is 
non-vanishing.\cite{Fu1,TI1,TI2}  These 
integer-valued topological invariants delineate TIs from conventional (`trivial')
insulators, with distinctive properties that are manifested primarily in edge 
states.\cite{NiuReview,TI1,Moore,Zhang}
Identification of TIs, and study of their properties and the critical point that
separates the phases, is highlighting new basic physics and several potential applications.

\begin{figure}[!htb]
\begin{center}
\includegraphics[width=0.32\textwidth,angle=0]{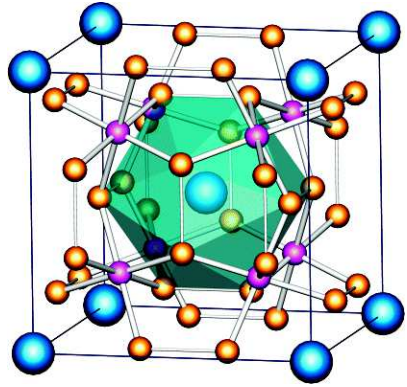}
\end{center}
\caption{(Color online)
Crystal structure of skutterudite minerals ($viz.$ CoSb$_3$) with
space group $Im\bar{3}$ (\#204), which includes inversion.  
The Co site (small pink [dark grey] sphere) is octahedrally coordinated to
Sb atoms (small yellow [light gray] spheres). Each Sb atom connects two octahedra,
as in the perovskite structure which has the same connectivity of octahedra.  The
large (blue) sphere denotes a large open site that is unoccupied in CoSb$_3$ but
is occupied in filled skutterudites (see for example Ref. [\onlinecite{maple}]).
The geometrical solid (center of figure) provides an indication of the volume
and shape of the empty region.
\label{Structure}
}
\end{figure}

Recently we reported\cite{OurPRL} on the skutterudite structure compound CoSb$_3$, which combines
and interrelates several of these properties (Dirac-Weyl semimetal at $k$=0; critical
point of degeneracy with massive bands; transition to topological insulator) 
in a unique way.  The skutterudite structure,
which is illustrated in Fig. \ref{Structure} and is discussed in Sec. II, is critical to
the electronic behavior in CoSb$_3$ though it may not be necessary for the unique type of
critical point that arises.  The relevant portion of the band structure, in a very
small volume of the Brillouin zone (BZ) centered at $\Gamma$, is unusually simple
compared to most of the TIs that have been discovered, $viz.$ the Bi$_2$Se$_3$ 
class,\cite{bi2se3}, the HgTe class,\cite{HgTe} and the three dimensional
system HgCr$_2$Se$_4$.\cite{GXu}

The electronic structure of skutterudites was studied early on by Singh and Pickett,\cite{djswep}
who focused on the peculiar valence band which was linear (except exceedingly near $\Gamma$)
and whose linearity extended surprisingly far out into the BZ.  It became clear that the
linearity of the band was responsible for the large thermopower\cite{caillat} that is
potentially useful in thermoelectric applications.\cite{tritt,sales} The origin of the
linearity was however not identified, 
but it leads to peculiar consequences:\cite{djswep} the density
of states varies as $\varepsilon^2$ near the band edge rather than
the usual three dimensional (3D) form $\sqrt{\varepsilon}$; as a consequence the
carrier density scales differently with Fermi energy $\varepsilon_F$;
the inverse mass tensor $\nabla \nabla \varepsilon_k$ is entirely
off-diagonal corresponding to an ``infinite'' transport mass; the
cyclotron mass differs from conventional 3D behavior, etc.
As the Sb sublattice is shifted in a symmetry-preserving way,\cite{OurPRL} the small
semiconducting gap of CoSb$_3$ closes at a critical point, giving rise to an unusual
occurrence: a Dirac-Weyl point in a 3D solid at the $\Gamma$ point. This transition 
point sets the stage for a conventional insulator to topological insulator transition,
although the specific behavior of the system at this point was not spelled out
in our earlier work. The only other topological insulator phases in skutterudite
materials that we are aware of are the examples by Jung {\it et al.}\cite{TIfilled}
in CeOs$_4$Pn$_{12}$ filled skutterudites.

This paper is organized as follows.  In Section II the skutterudite structure and its
relation to the symmetric perovskite (ReO$_3$) structure is reviewed. The computational methods
are outlined in Section III.  Section IV discusses the evolution of the band structure
during a perovskite-to-skutterudite transformation, including the development of the
gap and the appearance of the linear band.  Section V is devoted to constructing 
a microscopic but transparent understanding
of the electronic structure, including the development of the energy gap and the 
character of the peculiar linear band and the lower conduction band triplet that it
interacts with near, at, and beyond the critical point. In Section VI we provide 
details of the band inversion leading to the topological transition, and analyze
the anisotropy at the critical point. A brief summary is provided in Sec. VII.

\noindent

\section{Structure and Relation to Perovskite}
It is useful for the purposes of this paper to consider the skutterudite structure
to be a strongly distorted perovskite ATPn$_3$ structure in which the A atom is
missing ({\it i.e.} the ReO$_3$ structure) and the interconnected 
TPn$_6$ octahedra are rotated substantially (T = transition metal; Pn = pnictogen). The
skutterudite structure,\cite{struct1,struct2,struct3} pictured in Fig.~\ref{Structure}, has
space group $Im{\bar 3}$ (\#204) and a body-centered cubic (bcc) Bravais
lattice, and is comprised of a bcc repetition of four formula units (f.u.) when
expressed as $TPn_3$.  The pnictide ($Pn$) atoms form bonded units (nearly square but
commonly designated as ``rings'') which are not required
by local environment or overall symmetry to be truly square.
The three Pn$_4$ squares in the primitive
cell are oriented perpendicular to the coordinate axes.  Transition metal ($T$)
atoms (usually in the Co column) lie in six of the subcubes of the large cube of lattice
constant $a$; the other two subcubes (octants) are empty.  The structure
has inversion symmetry and is symmorphic, with 24 point group operations. 
The cubic operation that is missing is
reflection in (110) planes. The related 
filled skutterudites $XT_4$Pn$_{12}$ have an atom $X$ incorporated into the large
$2a$ site of $3\bar m$ symmetry\cite{harima} that remains empty in the compounds that we discuss.

\noindent
\begin{figure}[!htb]
\begin{center}
\includegraphics[width=0.32\textwidth,angle=0]{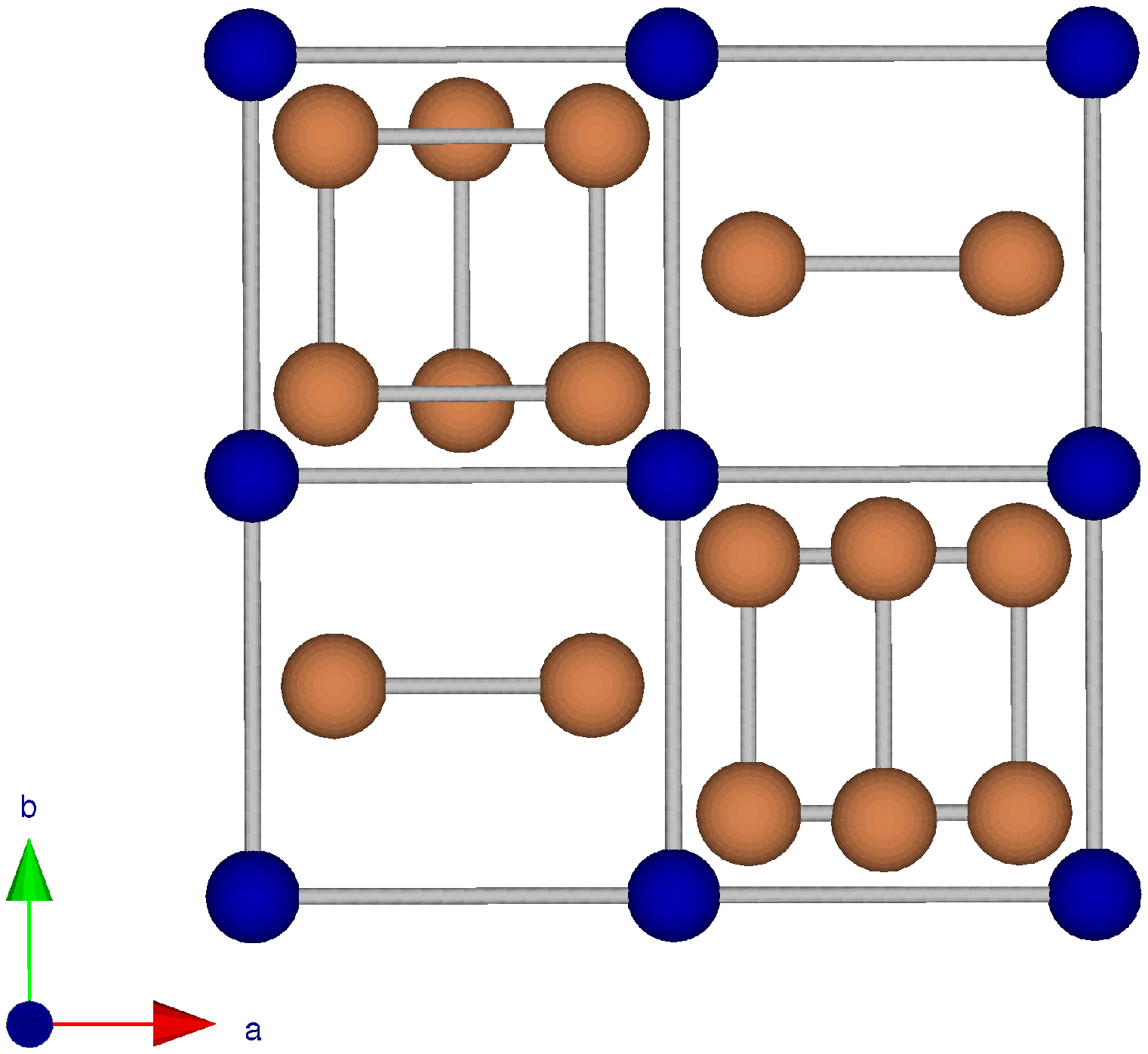}
\includegraphics[width=0.32\textwidth,angle=0]{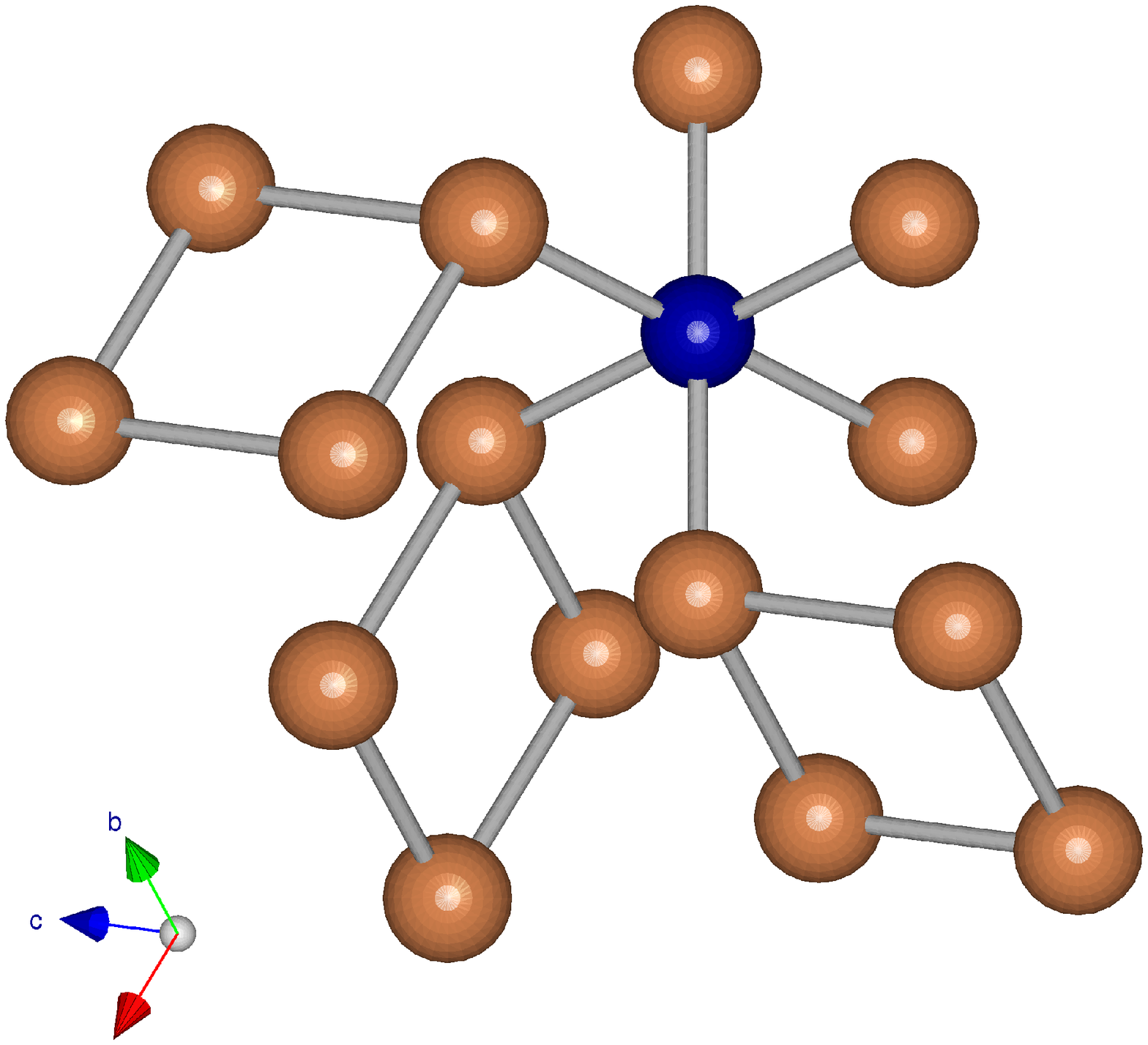}
\end{center}
\caption{(Color online)
Top: view down a cubic axis of CoSb$_3$, illustrating the very regular appearing
positions of the Sb$_4$ rings, drawn as connected by bonds. Note that one 
of these rings looks like a dumbbell
when viewed edge on. The (dark) Co atom is pictured without bonds.
Bottom: an indication of the environment of the Co atom (smaller sphere);
only three of the six attached Sb$_4$ rings are pictured.  In this structure,
all Sb sites are equivalent as are all Co sites. 
\label{SubStructure}
}
\end{figure}

Figure \ref{SubStructure} illustrates two aspects of the structure.  The top panel
reveals that, although the Sb$_6$ octahedra are rotated considerably (and strained)
compared to the perovskite structure, a great deal of regularity is retained. The
bottom panel gives a picture of the relationship between the Co atoms and the six
Sb$_4$ rings that it is coordinated with (only three mutually orthogonal rings are pictured). A natural
local coordinate system for describing an Sb$_4$ unit is to let it lie in the $x-y$
plane; then the $p_z$ orbitals we will discuss will be oriented perpendicular to
the plane of the Sb unit.  The $p_x, p_y$ orbitals are nearly symmetry related 
and naturally lead to bonding and antibonding molecular orbitals, and 
hence are very similar in character, while the nonbonding $p_z$ orbitals have
distinctive character.

If the origin is chosen so the large empty sites are centered at 
($\frac{1}{4},\frac{1}{4},\frac{1}{4}$)
and ($\frac{3}{4},\frac{3}{4},\frac{3}{4}$), then the rings are centered in each of
the other subcubes ($\frac{1}{4},\frac{3}{4},\frac{3}{4}$) etc.  A ring oriented
perpendicular to $\hat z$ is neighbored by rings above and below ($\pm \hat z$ directions) 
oriented perpendicular to (say) $\hat y$, by rings in the $\pm \hat x$ directions
perpendicular to $\hat x$, and neighbored by the empty sites along the $\pm \hat y$
directions. A symmetric combination of the four $p_z$ orbitals on a ring, call
it $P_z$,  is orthogonal to the $P_z$ orbitals on neighboring rings, so if 
dispersion is governed by inter-ring hopping (rather than through Co atoms)
some rather flat Sb $p_z$ bands should result.  Such behavior is seen in
fatbands plots (see below).

As mentioned, the skutterudite structure is related to the perovskite
structure $\Box T$Pn$_3$ ($\Box$ denotes an empty A site). Beginning from
perovskite, a rotation of the octahedra keeping the Pn atoms along the cube
faces results in the formation of the (nearly square) Pn$_4$ rings, and the
Pn octahedra become distorted and less identifiable as a structural feature. 
The transformation is, in terms of the internal coordinates $u$ and $v$,
\begin{eqnarray}
u'(s) = \frac{1}{4} + s (u - \frac{1}{4});~~
v'(s) = \frac{1}{4} + s (v - \frac{1}{4}).
\end{eqnarray}
The transformation path, from perovskite for $s$=0 to the observed structure
for $s$=1, is pictured in Fig. 1 of Ref. \onlinecite{Llunell}.
Below we make
use of this transformation to follow the opening of the (pseudo)gap between
occupied and unoccupied states.

\section{Computational Methods}
Two all-electron full-potential codes, FPLO-9\cite{fplo1} 
and WIEN2k\cite{wien2k} based on the augmented plane wave$+$local orbitals
(APW$+$lo) method,\cite{sjo} have been used in these calculations, with
consistent results. The Brillouin zone was sampled with regular 
$12\times 12\times 12$ $k$-mesh during self-consistency.
For WIEN2k, the basis size was determined by R$_{mt}$K$_{max}$= 7. 
Atomic radii used were 2.50 a.u. for Co and 2.23 a.u. for Sb, and the
Perdew-Wang form\cite{PW92} of exchange-correlation functional was used.
The experimental lattice parameters and atomic positions were taken from 
Ref. \onlinecite{struct1,struct2,struct3}.

\section{Perovskite to Skutterudite Transformation}
\begin{figure}[!]
\begin{center}
\rotatebox{-90}{\includegraphics[width=0.38\textwidth,angle=0]{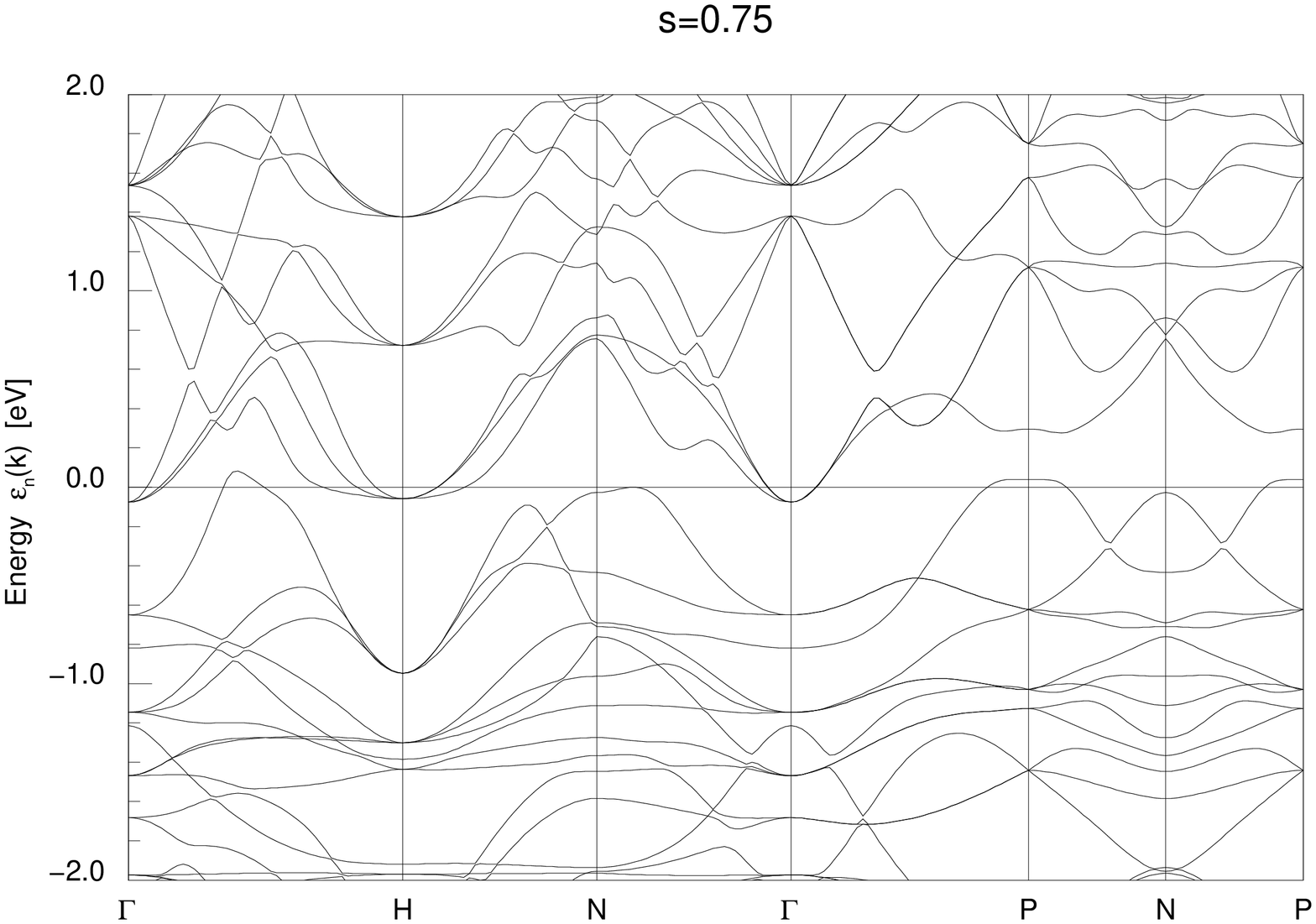}}
\rotatebox{-90}{\includegraphics[width=0.38\textwidth,angle=0]{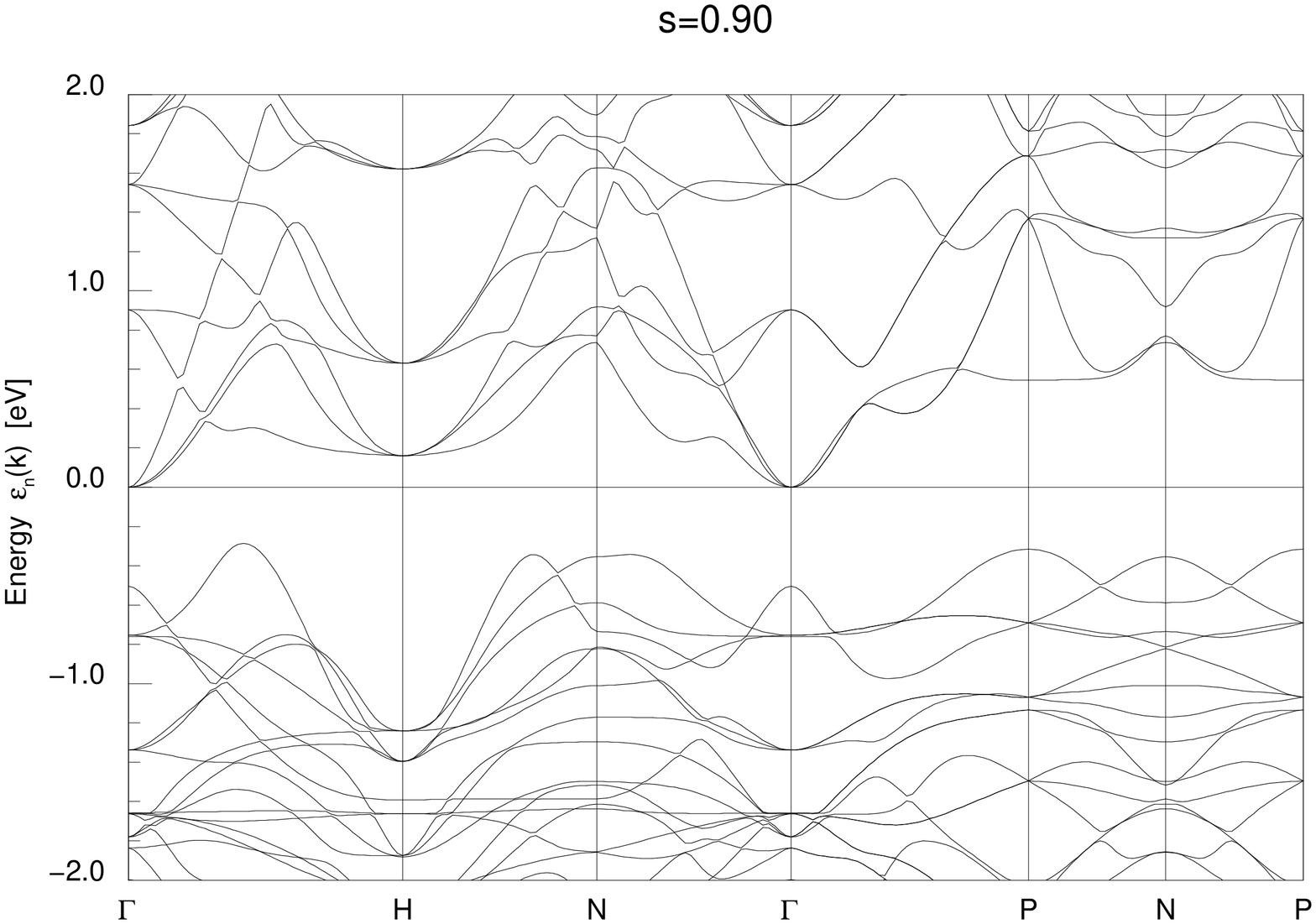}}
\rotatebox{-90}{\includegraphics[width=0.38\textwidth,angle=0]{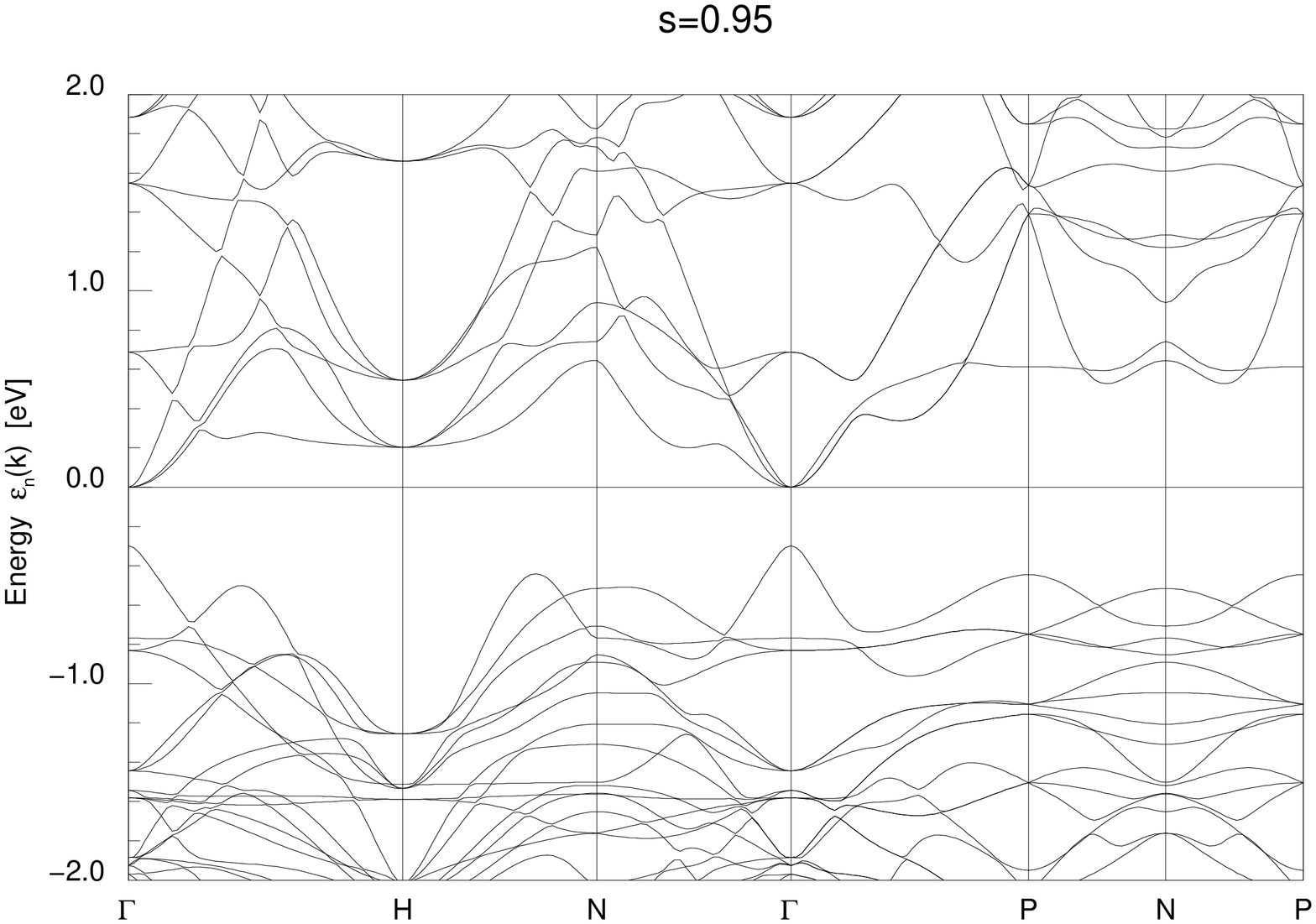}}
\end{center}
\caption{
Transformation of the band structure from metallic in the perovskite to
semiconducting in the skutterudite, following the (linear) transformation
described in the text. From the top, the panels show $s$ = 0.75, 0.90, 0.95.
The gap only begins to emerge around $s$= 0.75, where the conduction bands
are already disjoint from the valence bands. The linear band near $\Gamma$
emerges from the valence bands around $s$=0.95.
\label{fromPerov}
}
\end{figure}

The development of the electronic spectrum as the crystal is distorted
from perovskite to the skutterudite structure is pictured in the three
panels of Fig. \ref{fromPerov}.
For the ideal perovskite the result is a highly metallic state, with no gap
or pseudogap near the Fermi level.  The
important features only arise near the end of the distortion path. For
$s$=0.75 (top panel of Fig. \ref{fromPerov}) the valence bands have just become
disjoint from the conduction band and gap formation is imminent. For
$s$=0.90 the gap is well formed and the minimum direct gap is emerging
at the $\Gamma$ point. The unusual high velocity band arising from the
$\Gamma$ point, already striking at $s$=0.75, remains unchanged at
this point. The valence bands are a mass of indecipherable spaghetti.

By $s$=0.95 (bottom panel) a similarly high velocity band is emerging
from the valence bands to the maximum at the $\Gamma$ point. At $s$=1.0
it becomes clear that it is a partner of the high velocity conduction
band, as discussed in our earlier paper\cite{OurPRL} and to which
we return in the next section.  No doubt the formation of the gap,
{\it i.e.} formation of occupied bonding bands and unoccupied antibonding
bands, is behind the  stability of the skutterudite structure in the
Co pnictides class of materials.  The fact that the gap only opens as
the Sb $p_x$ and $p_y$ bonding-antibonding splitting becomes strong
supports the picture that this $t_{pp\sigma}$ interaction takes precedence
over the Co-Sb interactions within the CoSb$_6$ unit; however, these latter
interactions ({\it i.e.} the presence of the Co atom) are necessary for
the gap formation.

\noindent

\section{Electronic Structure}

\subsection{The Zintl Viewpoint}
Although in the `parent' perovskite system the structure is that of 
vertex-connected CoSb$_6$ octahedra, in the strongly distorted skutterudite
structure the Sb$_4$ rings form a basic structural motif.  
These compounds, viz. CoSb$_3$, are sometimes characterized as Zintl materials in
which the Sb$_4$ unit balances the charge of the Co unit.
Within the Zintl
picture, the charge of the Co$^{m+}$ ion must be balanced 
by a (Sb$_4$)$^{(4m/3)-}$ unit: there are only 3/4 as many Sb$_4$ rings as
Co atoms.  Thus there is tension between the Zintl picture, which does not
lead to integral formal charges on both of the primary units, and the 
presence of the gap that specifies an integral number of occupied bonding states.
Several papers\cite{djswep,jung,Llunell,Sofo,lefebvre,harima,koga,wei,lu,wee}
have presented results and some analysis for empty
and for a few filled skutterudites.  We have benefited from the previous work as
we proceed on a deeper analysis. 

We first follow the commonly held line of reasoning that the short Sb-Sb 
distance in the Sb$_4$ ring is
the most fundamental aspect of the electronic structure.\cite{jung}  The $p_x-p_x$ bonding 
along an $x$-axis
is characterized by a hopping amplitude $t_{pp\sigma} \approx$ 3 eV, so the
corresponding bonding and antibonding states lie at $\varepsilon_p \pm t_{pp\sigma}$.
The separation of Sb atoms and the $\sigma$ bonding and antibonding 
between $p_y$ orbitals along the
$y$-direction is indistinguishable from that of the $p_x$ $\sigma$ case.  
Indeed, the corresponding $p_{\sigma}$
bonding and antibonding character (not shown in figures) is centered roughly $\pm$3-4 eV 
below and above the
gap.  The on-site energies $\varepsilon_{p_x} \approx \varepsilon_{p_y}
= \varepsilon_p$ therefore lie
in the vicinity of the gap.  This picture leads to half-filled $p_x$ and $p_y$ orbitals. 

\noindent
\begin{figure}[!htb]
\begin{center}
\includegraphics[width=0.35\textwidth,angle=0]{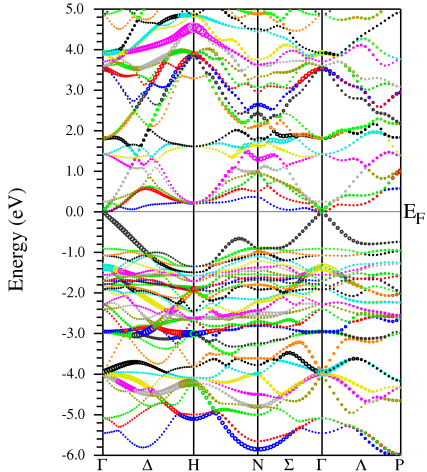}
\end{center}
\caption{(Color online)
Band structure of CoSb$_3$ along high symmetry directions, with fatbands
highlighting the Sb $p_z$ character.  The character is substantial 
in the -3 to -6 eV region, in the 3 to 5 eV region, but most notably in
the linear band extending through the gap, which is primarily $p_z$ in
character.
\label{pzBands}
}
\end{figure}

The $p_z$ orbitals within an Sb$_4$ unit are orthogonal to the $p_x, p_y$
orbitals and couple by a $t_{pp\pi}$ hopping amplitude. This coupling leads to
doubly degenerate levels at $\pm t_{pp\pi}$ relative to $\varepsilon_p$. 
The evidence is that there is negligible crystal field splitting 
of the three on-site $p$
orbitals. However, inter-unit $p_z-p_z$ interaction will be stronger than this
intra-unit coupling, and the $p_z$ orbitals also couple to the Co orbitals. The $p_z$
character, provided by the fatbands representation in Fig. \ref{pzBands}, is in
fact distributed throughout the -6 eV to +5 eV region and is not reproducible by
a simple model due to other couplings.  The $p_z$ projected density of 
states (DOS) indicates the
$p_z$ orbitals are at least half-filled, perhaps slightly more.

\noindent
\begin{figure}[!htb]
\begin{center}
\includegraphics[width=0.35\textwidth,angle=0]{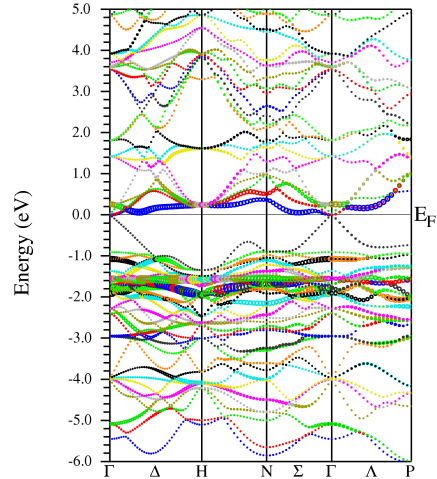}
\end{center}
\caption{(Color online)
Band structure of CoSb$_3$ along high symmetry directions, with fatbands
highlighting the Co $3d$ character. This plot reflects a narrow mass of
$d$ bands throughout the -1 to -2 eV range, more around -3 eV, and $\sim$4
unoccupied bands in the 0-1 eV range.  The amount of $3d$ occupation is
discussed in the text.
\label{CoBands}
}
\end{figure}

\subsection{Complications with the Zintl Viewpoint}
It is readily demonstrated that analysis focusing primarily on the
Sb$_4$ rings has strong limitations.  We have performed calculations
with the Co atom removed, $\Box$Sb$_3$.  The Sb $p$ band complex is
broad (10 eV) and featureless.  There is no hint of a gap
or even pseudogap, indicating that the $p_x$ and $p_y$ bonding-antibonding
feature described above is far from a dominant feature. Only when Co is re-inserted
into the structure does the remarkably clean 1 eV gap become
carved out around the Fermi level, except of course for the peculiar linear band
near the zone center.  The gap
therefore arises from the Co $3d$ mixing (evidently strongly) with Sb $5p$ orbitals,
with the linear band emerging from the valence bands being of strong Sb
character.

Figure \ref{CoBands} illustrates the Co $3d$ character.  It lies
mostly below the gap, in the very narrow -1 to -2 eV range.  A much
smaller amount of character, roughly something like four bands, lies immediately
above the gap in the 0-1 eV range.  These two parts of the $3d$ projected
DOS are neatly and impressively split by the 1 eV gap.  The ``charge state''
(or formal valence, or oxidation state)
of the Co atom is a question that can be asked, since this is an insulating
material.  If there are four unoccupied
$d$ bands (though this cannot be claimed very conclusively due to the strong
hybridization) then considering
there are four Co atoms in the cell, the occupation would be $d^8$ Co$^{1+}$.
A nonmagnetic ion with this filling is natural for the Co $\bar{3}$ site symmetry.
As the CoSb$_6$ octahedron is distorted strongly in progressing from the cubic
perovskite structure to the skutterudite one, the crystal field levels reduce as
$t_{2g} \rightarrow t_1 + t_2$, the latter being the doublet, and $e_g\rightarrow
e_1 + e_2$.  One unoccupied orbital singlet (both spins) gives the $d^8$ occupation.
The magnetic quantum number $m=0$ orbital with respect to the local three-fold axis
is the natural one for the $3d$ holes to occupy. 

Charge balance will then leave the rings as (Sb$_4$)$^{-4/3}$, a very unsatisfactory
result for accounting for the gap: each Sb$_4$ ring would not have an
integer number of (fully) occupied bands (or molecular orbitals).  Within the
picture that the $p_x$ and $p_y$ orbitals are half-filled due to the large
$t_{pp\sigma}$ bonding-antibonding splitting, this Co charge state would
suggest $p_z$ orbitals to be 1/6 more than half-filled, a peculiar result and one
that would only be compatible with a metallic, rather than semiconducting, state.  

The integrated DOS for the Co $3d$ states in fact gives a clear $d^8$ occupation.
However, with near-neutral atoms there might be significant Co $4s, 4p$ occupation,
which could bring Co nearer to a neutral $d^8s^1$ configuration.  Some
Co $sp$ character is in fact found in the upper valence bands, though it is possible
that this represents tails of Sb $p$ orbitals.  The Co $d$ occupation $d^8$ is clear,
however.

\subsection{Sb $6p$ - Co $3d$ Mixing}
It was noted above that when Co is removed from the structure, there is no
gap nor any indication of one, nor any candidate for a potential linear band.  
The most direct interaction of the Sb $p_z$ orbitals on
the ring is between the $p_z$ orbital on the nearest Sb in the ring with the 
rotated $xz \pm yz$ orbitals on the eight Co neighbors.  Sb$_6$ molecular orbitals
(MOs) formed by linear combinations of $p_z$ orbitals can be sorted according to whether
they are even or odd parity with respect to the center of inversion at the Co site.
The even parity MOs mix with the Co $3d$ orbitals, forming bonding and antibonding
pairs tending to open a gap; the odd parity MOs do not mix with the $3d$ orbitals.  
As a result of this coupling, four of the five $3d$ orbitals are 
lowered and occupied, while one
is raised and unoccupied, giving the 1 eV gap and the $d^8$ occupation.

\subsection{Band Character near the Gap}
The fatbands representation of Sb $p_z$ character in Fig. \ref{pzBands} 
shows the substantial $p_z$ character of the linear band in the upper
valence bands near the zone center.  This $p_z$ character is in fact
the dominant character, although various other small contributions from
other valence orbitals are mixed in by the low Sb site symmetry.
The character of the 3-fold degenerate conduction band minimum state is
more complex. These bands have primarily Co ``$t_{2g}$'' character; the quotes
arise from the fact that the degeneracies of the cubic
$t_{2g}$ and $e_g$ subshells are broken
by the distortion of the octahedra in the skutterudite structure. 
There is a strongly
$d_{z^2}$ flat band just above (+0.2 eV) the gap, and additional strong
$d_{x^2-y^2}$ character in the 0.2-0.6 eV region. 

The band figures we show below demonstrate that the linear band couples
to only one of the triplet of conduction bands at and near $\Gamma$,
{\it i.e.} couples to only one linear combination of the triplet states.
It does not couple to the Co $3d$ orbitals.

\noindent
\begin{figure}[!htb]
\begin{center}
\rotatebox{-00}{\includegraphics[width=0.35\textwidth,angle=0]{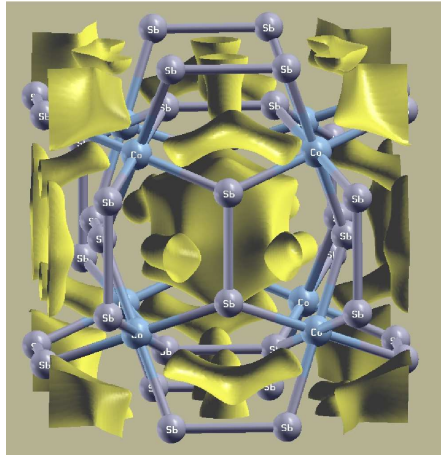}}
\end{center}
\caption{
Surface plot of the density arising from the linear band in CoSb$_3$ near
$k$=0. The closed surface surrounds the unoccupied site in the skutterudite
lattice and encloses a region of {\it lower density} than outside of the
surface.
\label{density}
}
\end{figure}

\subsection{A consistent viewpoint: an empty-site molecular orbital}
A generalized viewpoint of the most physical orbitals in CoSb$_3$ is that 
of four Co$^{1+}$ $d^8$ atoms (as discussed above), three Sb$_4$
rings with $p_x$- and $p_y$-$\sigma$ bonding states filled and antibonding states
unoccupied, and molecular orbitals formed from the twelve $p_z$ orbitals and
centered on the empty $2a$ site.  This site is an inversion center, so the
MOs can be classified as parity-even or -odd. (Note that these are different
MOs than were discussed in Sec. V.C.) Twelve MOs can be formed,
which can be classified also by the tetrahedral $2a$ site symmetry. Charge counting
requires that eight of these MOs must be occupied (with both spins).

We will not attempt to identify the positions and dispersions of these MOs,
except to note that linear band is one of these MOs.  Jung {\it et al.}\cite{jung}
identified this MO as $A_{2u}$, an orbitally nondegenerate state 
with odd parity at $k$=0, 
and therefore not mixing with the Co $3d$ orbitals at (or near) $\Gamma$. 
The density arising from this orbital is of interest. We had earlier investigated
the possibility that an electron may reside in the large empty $2a$ site in the
skutterudite structure.
This type of ``electride'' configuration, in which an interstitial electron without
a nucleus becomes an anion, is well established in molecular
solids.\cite{crownether,dye} However, the density at the $2a$ site is very small,
ruling out this possibility.

An isosurface plot of the density arising from the linear band near $\Gamma$ is
shown in Fig. \ref{density}. The closed surface in the density encloses a region
of {\it low density} centered on the unoccupied site. Exploring the density with
isosurfaces at various values of density reveals no local maximum at this large
interstitial site, thus no electride-like character.  
Lefebvre-Devos {\it et al.} have presented\cite{lefebvre} complementary isosurface
density plots of this band, with the high density regions of two types.  One region
surrounds the empty site (with rather complex shape) consistent with its origin
as a MO comprised of $p_z$ Sb$_4$ ring states, with the tetrahedral symmetry of the $2a$ site. 
The second region is within each CoSb$_6$ octahedron, centered on either side of the
Co atom along the local $\bar{3}$ axis, appearing as a three-bladed propeller that
arises from the lobes of the $p_z$ orbitals that lie closer to the Co atom.

Interaction of the $p_z$ states with Co $3d$ orbitals appears to cause 
a majority of the density of this state to border the large empty site. The result
is not an electride state but rather a large molecular orbital (one of twelve in
total) centered on the
empty site, with one of them ($A_{2u}$) per primitive cell.  We have previously reported
that this state at $\Gamma$ is parity-odd.\cite{OurPRL} 

Such an orbital has the same hopping amplitude $t$ along each nearest neighbor
direction on a $bcc$ lattice, hence its dispersion looks like that of an
$s$ orbital, $E_k$ = 8$t$ cos($k_xa/2$) cos($k_ya/2$) cos($k_za/2$), and does
not lead to a linear band at $k$=0.  This shouldn't be surprising, as the
true linearity at $\Gamma$ is accidental, due to a degeneracy that has to be
tuned. In our previous paper,\cite{OurPRL} this tuning was accomplished by
making two on-site energies in the tight-binding model identical.  

The ``linear band'' aspect
of the skutterudites may have been over-interpreted, because it arose in the
most common member (the mineral skutterudite CoAs$_3$) which was studied
most heavily initially.  In a general skutterudite, for example the filled
one LaFe$_4$P$_{12}$, the same band rises out of the valence band complex but
is not unduly ``linear.''  The isovalent members XPn$_3$, with X = Co, Rh,
Ir, and pnictides Pn = As and Sb, all seem to lie not far from the critical
degeneracy that extends the band linearly to $k$=0; however, we know of nothing
in general about the skutterudite crystal structure that otherwise promotes
a linear band.


\section{Nonanalyticity, Anisotropy, and the Topological Insulator Phase}
\subsection{Band Inversion: Internal strain, spin-orbit coupling, and
    tetragonal strain}
In previous work\cite{OurPRL} we demonstrated that it is possible, with very small Sb sublattice
displacement (see Sec. II for the linear distortion, characterized by the parameter
$s$), for this band to remain linear all the way
to $k$=0, marking a critical point.  In that work, the progression of the band crossing
at the critical point was plotted. Due to the 3-fold degeneracy of the conducting band
edge, the progression versus $s$ was from semiconductor for $s\leq 1.19$, to the 
critical point at $s$=1.020 with linear {\it valence and conduction} bands degenerate
with two quadratic bands, to a zero-gap semiconductor for $s\geq 1.021$ due to a
symmetry-determined degeneracy, even when spin-orbit coupling (SOC) is included. 
Due to the band inversion and the character of the states at $\Gamma$, the transition
is from trivial insulator to one with topological character. Because the degeneracy
at the point Fermi surface (at $\Gamma$) remains, the latter state is a topological 
zero-gap semiconductor rather than a topological insulator.

\noindent
\begin{figure}[!htb]
\begin{center}
\rotatebox{-00}{\includegraphics[width=0.35\textwidth,angle=0]{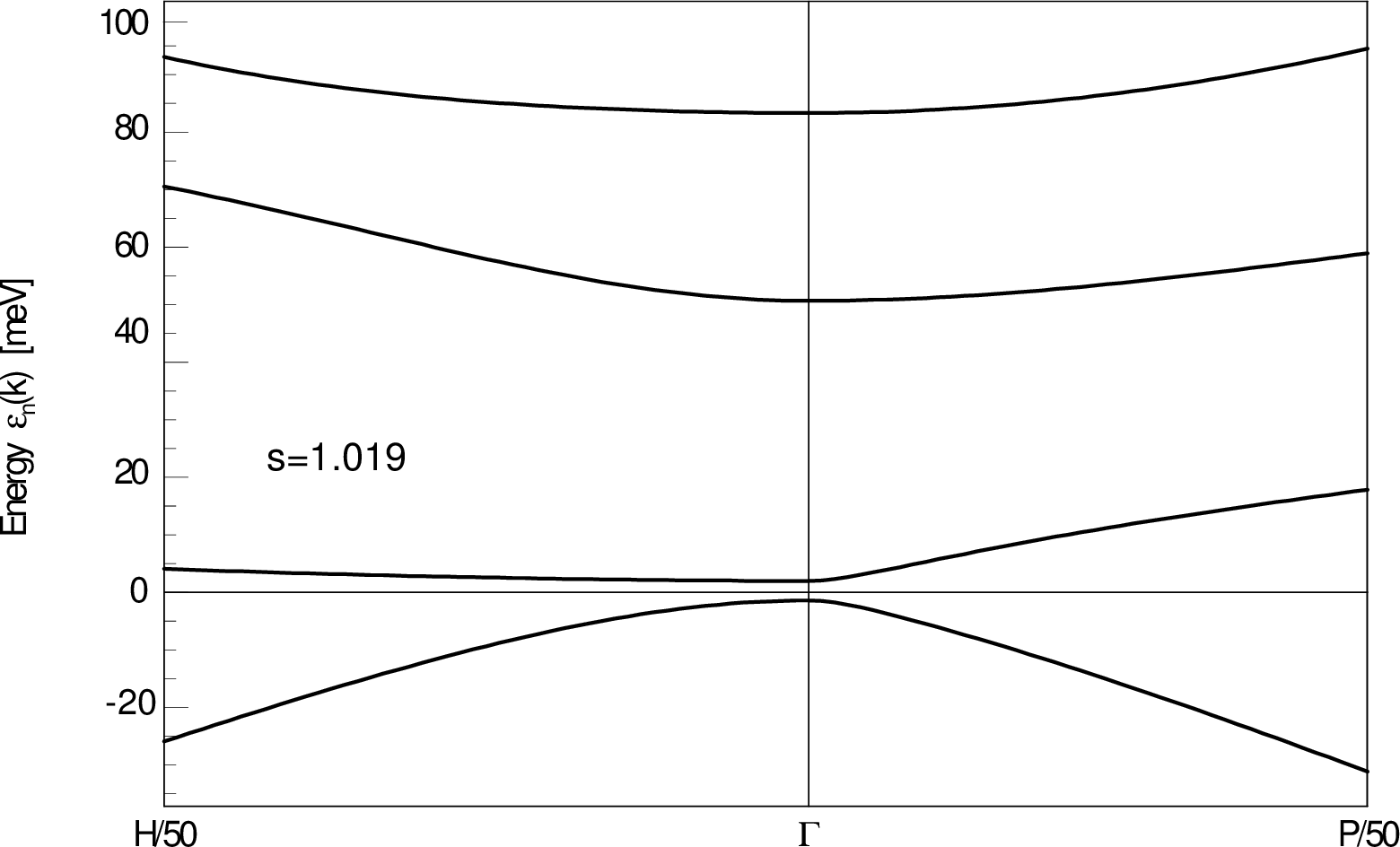}}
\rotatebox{-00}{\includegraphics[width=0.35\textwidth,angle=0]{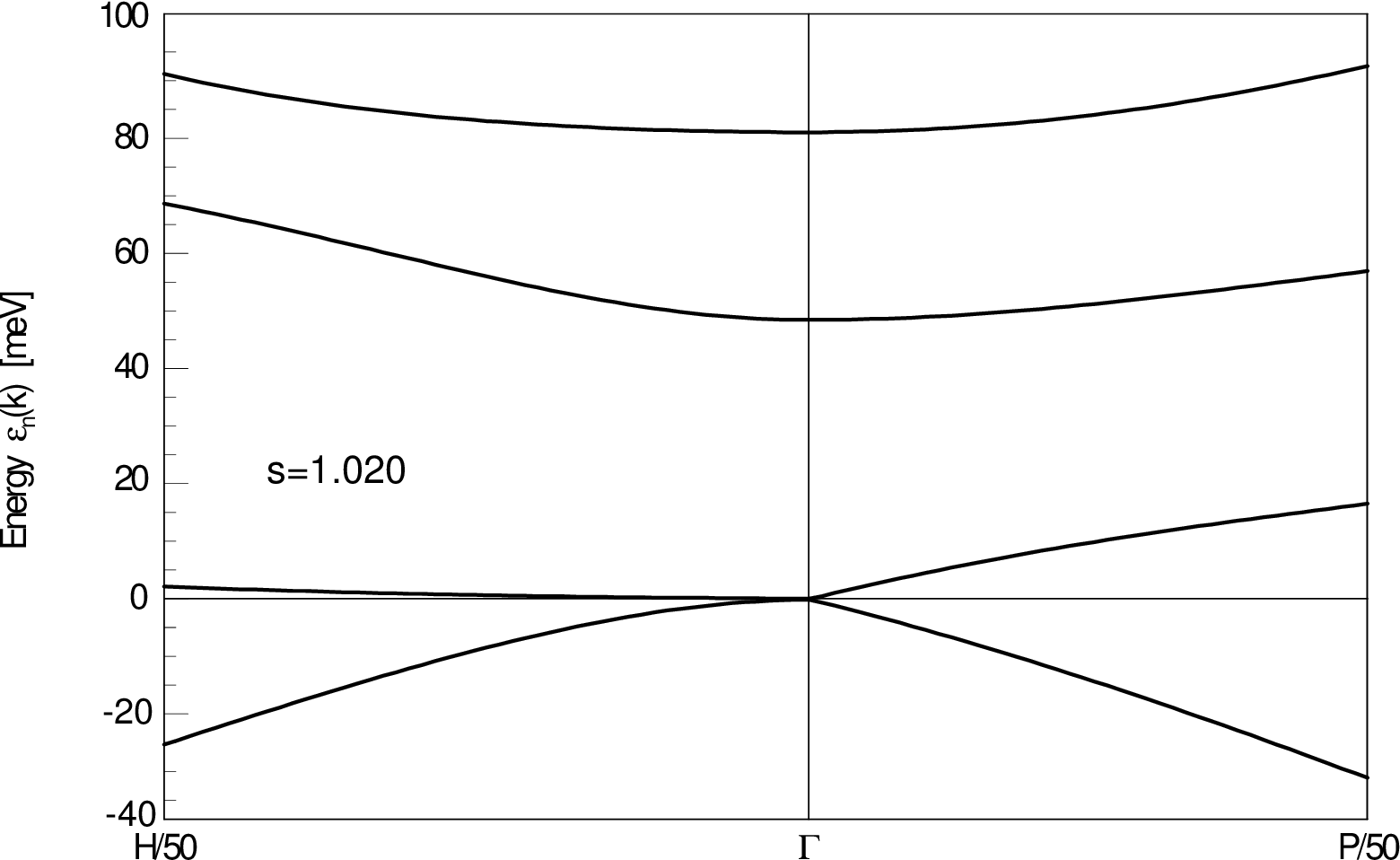}}
\rotatebox{-00}{\includegraphics[width=0.35\textwidth,angle=0]{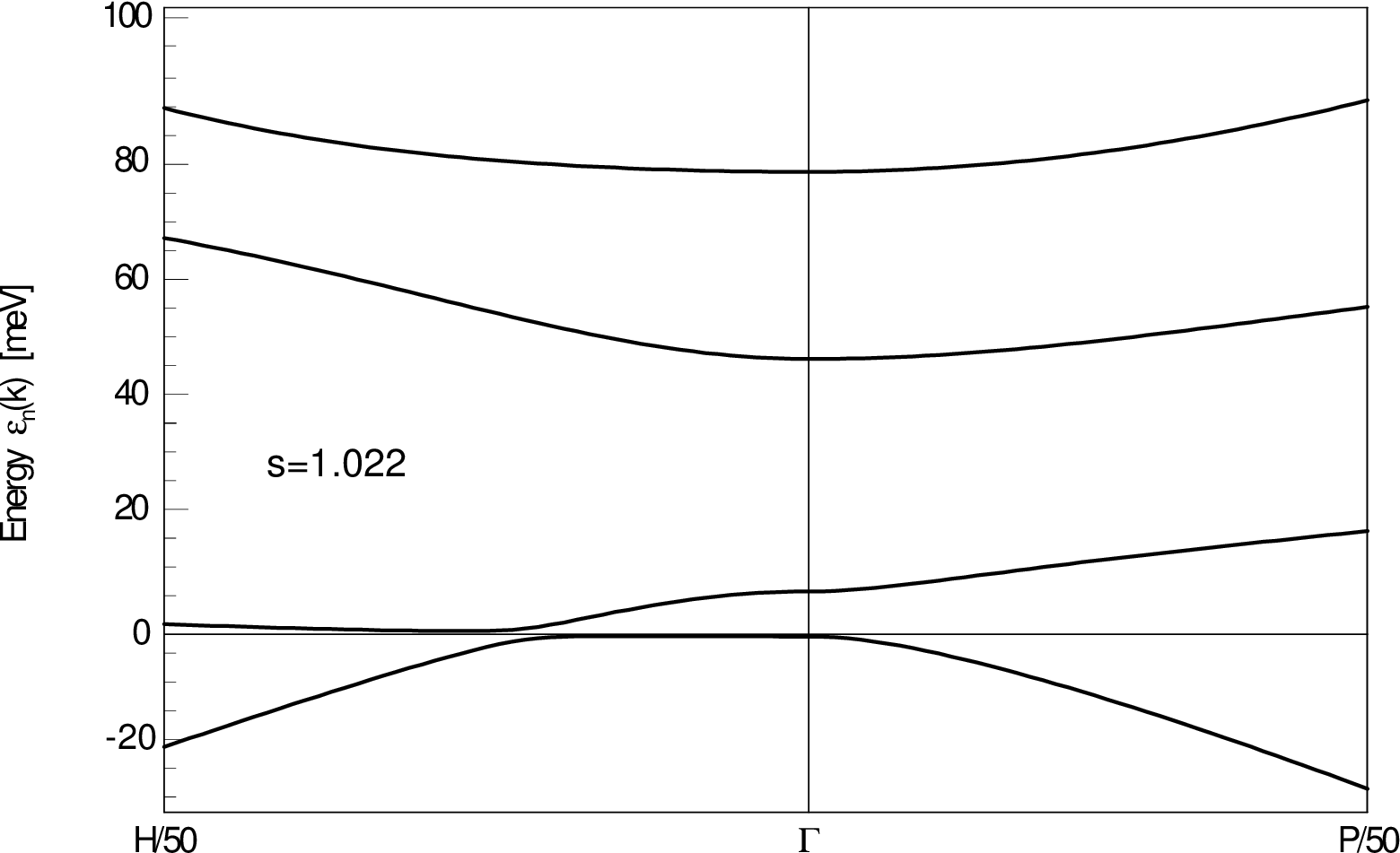}}
\end{center}
\caption{
Transformation of the band structure very near $k$=0 from insulating at $s$=1.019
(top) through the semimetallic critical point at $s$=1.020 to the inverted band
insulator at $s$=1.022 (bottom). H/50 and P/50 denotes the point 2\% of the
way from $\Gamma$ to H and P, respectively. The horizontal line lies in the gap (or at the
Fermi level) in each panel.  In the bottom panel the gap has moved out along
the $\Gamma$-H line but is non-vanishing. The much larger band repulsion along
$\Gamma$-P produces the unique band dispersion at the critical point.
\label{transform}
}
\end{figure}

We now demonstrate that a simple tetragonal distortion, such as might result from growth
on a substrate with small lattice mismatch, does produce a topological insulator (TI)
but with an unexpected placement of the gap.  The progression of the band structure
very near $\Gamma$ is shown in Figure \ref{transform} for strain $c/a$=1.01. 
The gap before the critical point is reached always
has quadratic bands for sufficiently small $k$. At the critical point $s= 1.020$ (when
the strain is imposed) leads, unexpectedly, to linear bands only in certain of the
high symmetry directions (such as the $<111>$ direction, toward the P point). In other
high symmetry directions (from $\Gamma$ toward H is pictured) the bands are quadratic
and the upper one is quite flat (heavy mass). We return to this point in the next
subsection.

Beyond the critical point, the four bands become non-degenerate once more; the
lower (linear) band has crossed the lowest of the three conduction bands and
acquired mixed character.  The coupling
between bands is clearly much smaller along the axes ($\Gamma$-H direction) than 
along the other symmetry lines, and the gap occurs along this direction.  Because
of the heavy mass of the lower conduction bands, the gap is quite small, being of
the order of 1 meV for the tetragonal strain that is pictured in Fig. \ref{transform}.

\subsection{Anisotropy at the Critical Point}
Inspecting the bands along the three cubic high symmetry directions, a second peculiar
feature becomes evident.  At the critical point, the ``linear band'' disperses linearly
along the $<111>$ directions, while they are quadratic (with distinct masses)
along both the $<100>$ and $<110>$ directions.  The directional dependence as well as
the magnitude dependence of the wavevector is anomalous.  The form of the dispersion
for small $|\vec k|$ can be constructed.  A polynomial in the direction of $\vec k$
($\hat k$) with cubic symmetry that vanishes along the $<100>$ directions can be
constructed as
\begin{eqnarray}
P_{100}(\hat k)&=& \sqrt{\prod_{j=1}^6 (\hat k -\hat k_j)^2} \nonumber \\
               &=& \prod_{j=1}^6 |\hat k - \hat k_j|,
\end{eqnarray}
where \{$\hat k_j$\} are the six $<100>$ unit vectors. The denominator normalizes the
polynomial to unity along each of the $<111>$ directions; $\hat k_{111}$ lies along
the $<111>$ direction (any of them).   The analogous polynomial $P_{110}(\hat k)$
that vanishes along each $<110>$ direction and is normalized to unity along each $<111>$
direction can be constructed.  Then the peculiar band pair is given by
\begin{eqnarray}
\varepsilon(\vec k) = \pm v |\vec k| P_{100}(\hat k) P_{110}(\hat k) 
   + \frac{k^2}{2m_s} \pm \frac{k^2}{2m_a},
\end{eqnarray}
where $m_s, m_a$ serve as symmetric and antisymmetric masses.

\subsection{Isovalent Skutterudite Antimonides}
We have studied the band structures of antimonide sisters of 
CoSb$_3$, based on the transition metal atoms  Rh and Ir.
Band structures, or sometimes only gap values, of these compounds have been 
reported earlier.\cite{harima,takeg,koga} 
The band structures around the gap at $k$=0 differ somewhat in the literature, due
to sensitivity to structure (and that some use relaxed coordinates while others
use the measured structure), to the level of convergence of the calculation, and
the exchange-correlation functional.

The difference in lattice constant is compensated very closely 
by the effect of difference in
interatomic distances on hopping amplitudes, and the resulting bands are very similar.
The single feature of interest is the
magnitude of the gap (and whether positive or negative, the latter giving an inverted
band structure) and the strength of SOC effects. As mentioned above, band inversion
alone does not in itself give rise to a topological insulator state in these compounds
because even after SOC is included, a two-fold degeneracy at $\Gamma$ is lowest and the
system is a point Fermi surface, zero-gap semiconductor. Such cases are 
referred to as topological semimetals.
However as shown above, the topological
semimetal can be driven into the TI phase by tetragonal
distortion that lifts the final degeneracy. 

{\it RhSb$_3$.} For a range of volumes around the experimental one ($a$= 9.23 \AA), 
this compound has an inverted band structure, hence it is a point Fermi 
surface, zero-gap topological semimetal. 
In the inverted band structure before considering SOC, 
the singlet lies 23 meV above the triplet 
that pins the Fermi level. Inclusion of SOC leaves the Fermi energy pinned to a doublet.
A tetragonal distortion raises the degeneracy;  a topological
insulator is obtained for $c/a > 1$ in the same way that happens for CoSb$_3$
although no sublattice strain is required in RhSb$_3$.

Generally, reducing volume (applying pressure) in these Sb-based skutterudites 
leads to a larger gap. In the case of RhSb$_3$, our calculations indicate a gap never 
actually opens within cubic symmetry, for peripheral reasons. When the 
band inversion at $\Gamma$ is undone and a gap at $\Gamma$ appears 
at sufficiently high pressure (a= 9.06 \AA), the material 
is metallic due to other (Rh $4d$) bands having become lowered in energy,
crossing the Fermi level along the $\Gamma$-N
direction.

{\it IrSb$_3$}. The larger SOC for heavier atoms led us to consider this $5d$ compound.
However, the linear band and several others have practically no metal atom
character and are not affected by its spin-orbit strength.
We obtain, using the same computational methods as for the other two compounds, that
like CoSb$_3$, IrSb$_3$ has a $\sim$80 meV gap and therefore no topological character at
the experimental equilibrium structure. Like CoSb$_3$, it can be driven to an
inverted band structure and hence topological
character by internal and tetragonal strains.

\section{Summary}
One goal of this work was to provide a simple but faithful picture of the electronic
structure of CoSb$_3$: the origin of the gap and the character of the linear band
being the most basic.  The picture is this.  Co is in a $d^8$ configuration, leaving
four electrons in the Co$_4$Sb$_{12}$ unit cell to go into other bands.  The $p_x$ and
$p_y$ orbitals on the Sb$_4$ ring form strongly bonding (occupied) and antibonding
(unoccupied) states, accounting for two of the three $5p$ electrons of each Sb
atom, for a total of twelve occupied $\sigma$-bonding states (of each spin). The
twelve $\pi$-oriented $p_z$ orbitals couple between units as well as with the Co $3d$
orbitals, and form ``molecular orbitals'' centered on the vacant $2a$ sites in the
skutterudites lattice. Eight of these MOs are occupied by 16 electrons, one from
each Sb and one from each Co.  The linear band is best pictured as arising from the
coupling and dispersion of these MOs.

We have also justified our earlier statement\cite{OurPRL} that CoSb$_3$ is very near 
a conventional-to-topological transition.  A small symmetry-preserving internal
strain, a small applied tetragonal strain, and spin-orbit coupling drives CoSb$_3$
into a topological insulator phase. This critical point also marks a point of
accidental (but tunable, by the strains) degeneracy that gives rise to emergence
of linear Dirac-Weyl bands emanating from $k$=0, where they are degenerate with
massive bands.  At this critical point, the dispersion is non-analytic at $k$=0
and anisotropy is extreme, with the mass varying from normal to vanishing. In
this respect CoSb$_3$ at its critical point bears several similarities to the
semi-Dirac (Dirac-Weyl) behavior encountered\cite{sD1,sD2,sD3} 
in ultrathin layers of VO$_2$.

\section{Acknowledgments}
We acknowledge illuminating discussions with S. Banerjee, R. R. P. Singh,
C. Felser, and L. M\"uchler.  
Work at UC Davis was supported
by DOE Grant DE-FG02-04ER46111. 
V.P. acknowledges support from the Spanish Government through
the Ram\'{o}n y Cajal Program.



\begin{thebibliography}{10}

\bibitem{book}I. M. Tsidilkovski, {\it Gapless Semiconductors: a New Class of Materials}
   (Akademie-Verlag, Berlin, 1988).
\bibitem{graph1} A. K. Geim and K. S. Novoselov, Nature Mater. {\bf 6}, 183 (2007).
\bibitem{graph2}A. H. Castro Neto, F. Guinea,
  N. M. R. Peres, K. S. Novoselov, and A. K. Geim,
  Rev. Mod. Phys. {\bf 81}, 109 (2009).

\bibitem{Fu1}L. Fu, C. L. Kane, and E. J. Mele, Phys. Rev. Lett. {\bf 98}, 106803 (2007).
\bibitem{TI1}M. Z. Hasan and C. L. Kane,
Rev. Mod. Phys. {\bf 82}, 3045 (2010).
\bibitem{TI2}X.-L. Qi and S.-C. Zhang,
Rev. Mod. Phys. {\bf 83}, 1057 (2011).


\bibitem{NiuReview}D. Xiao, M.-C. Chang, and Q. Niu, Rev. Mod. Phys. {\bf 82},
  1959 (2010).
\bibitem{Moore}J. E. Moore, Nature {\bf 464}, 194 (2010).
\bibitem{Zhang}X.-L. Qi, T. L. Hughes, and S.-C. Zhang, Phys. Rev. B {\bf 78}, 195424 (2008).


\bibitem{OurPRL} J. C. Smith, S. Banerjee, V. Pardo, and W. E. Pickett,
  Phys. Rev. Lett. {\bf 106}, 056401 (2011).

\bibitem{bi2se3}P. Larson, V. A. Greanya, W. C. Tonjes, R. Liu, S. D. Mahanti,
  and C. G. Olson, Phys. Rev. B {\bf 65}, 085108 (2002);
  C.-X. Liu, X.-L. Qi, H. J. Zhang, X. Dai, Z. Fang, and S.-C. Zhang,
  Phys. Rev. B {\bf 82}, 045122 (2010).

\bibitem{HgTe}C. Liu, T. L. Hughes, X.-L. Qi, K. Wang, and S.-C. Zhang, Phys. Rev.
  Lett. {\bf 100}, 236601 (2008).
\bibitem{GXu}G. Xu, H. Weng, Z. Wang, X. Dai, and Z. Fang, Phys. Rev. Lett.
   {\bf 107}, 186806 (2011).

\bibitem{djswep}D. J. Singh and W. E. Pickett, Phys. Rev. B {\bf 50},
   11235(R) (1994).

\bibitem{caillat}T. Caillat, A. Borshchevsky, and J.-P. Fleurial,
  J. Appl. Phys. {\bf 80}, 4442 (1996).

\bibitem{tritt} G. S. Nolas, D. T. Morelli, and T. M. Tritt, Annu. Rev. Mater. Sci.
   {\bf 29}, 89 (1999).
\bibitem{sales}B. C. Sales, D. Mandrus, and R. K. Williams, Science {\bf 272}, 1325 (1996).

\bibitem{TIfilled}B. Yan, L. M\"uchler, X.-L. Qi, S.-C. Zhang, and C. Felser,
   Phys. Rev. B (in press); arXiv:1104.0641.




\bibitem{maple}P.-C. Ho, N. A. Frederick, V. S. Zapf, E. D. Bauer, T. D. Do,
    M. B. Maple, A. D. Christianson and A. H. Lacerda,  
  Phys. Rev. B {\bf 67}, 180508(R) (2003).

\bibitem{struct1} K. Koga, K. Akai, K. Oshiro, and M. Matsuura, Phys. Rrev. B 
   {\bf 71}, 155119 (2005).
\bibitem{struct2}T. Schmidt, G. Gliche, and H. D. Lutz, Acta Crystallogr., Sect. C: 
    Cryst. Struct. Commun. {\bf 43}, 1678 (1987).
\bibitem{struct3}A. Kjekshus and T. Rakke,
    Acta Chemica Scand. {\bf 28A}, 99 (1978).



\bibitem{harima}H. Harima and K. Takegahara, J. Phys.: Condens. Matter
  {\bf 15}, S2081 (2003).

\bibitem{Llunell}
M. Llunell, P. Alemany, S. Alvarez, V. P. Zhukov, and A. Vernes,
   Phys. Rev. B {\bf 53}, 10605 (1996).


\bibitem{fplo1} K. Koepernik and H. Eschrig, Phys. Rev. B {\bf 59}, 1743 (1999).

\bibitem{wien2k} K. Schwarz and P. Blaha,
  Comp. Mat. Sci. {\bf 28}, 259 (2003).

\bibitem{sjo} E. Sj{\"o}stedt, L. N{\"o}rdstrom, and D. J. Singh,
  Solid State Commun. \textbf{114}, 15 (2000).
  
\bibitem{PW92}J. P. Perdew and Y. Wang, Phys. Rev. B {\bf 45}, 13244 (1992).


\bibitem{Sofo}J. O. Sofo and G. D. Mahan, Phys. Rev. B {\bf 58}, 15620 (1998).

\bibitem{lefebvre}I. Lefebvre-Devos, M. Lassalle, X. Wallart, J. Olivier-Fourcade, L. Monconduit, and J. C. Jumas, Phys. Rev. B
  {\bf 63}, 125110 (2001).


\bibitem{koga}K. Koga, K. Akai, K. Oshiro, and M. Matsuura, Phys. Rev.
  B {\bf 71}, 155119 (2005).

\bibitem{wei}W. Wei, Z. Y. Wang, L. L. Wang, H. J. Liu, R. Xiong, J. Shi,
  H. Li, and X. F. Tang, J. Phys. D: Appl. Phys. {\bf 42}, 115403 (2009).

\bibitem{lu}P.-X. Lu, Q.-H. Ma, Y. Li, and X. Hu, J. Magn. Magn. Mat.
  {\bf 322}, 3080 (2010).

\bibitem{wee}D. Wee, B. Kozinsky, N. Marzari, and M. Fornari, Phys. Rev.
  B {\bf 81}, 045204 (2010).


\bibitem{jung} D. Jung, M.-H. Whangbo, and S. Alvarez, Inorg. Chem. {\bf 29},
   2252 (1990).

  
\bibitem{crownether}D. J. Singh, H. Krakauer, C. Haas, and W. E. Pickett,
  Nature {\bf 365}, 39  (1993).
\bibitem{dye}J. L. Dye, Science {\bf 247}, 663 (1990).

\bibitem{takeg}K. Takegahara and H. Harima, Physica B {\bf 328}, 74 (2003).


\bibitem{sD1}V. Pardo and W. E. Pickett,  
Phys. Rev. Lett. {\bf 102}, 166803 (2009).
\bibitem{sD2}S. Banerjee, R. R. P. Singh, V. Pardo, and W. E. Pickett,  
Phys. Rev. Lett. {\bf 103}, 016402 (2009).
\bibitem{sD3}V. Pardo and W. E. Pickett,  
Phys. Rev. B {\bf 81}, 035111 (2010).



\end{thebibliography}
\end{document}